\documentstyle[pre,aps,psfig]{revtex}
\draft

\begin{document}                % INITIALIZE - DONT CHANGE

\def\be{\begin{equation}}
\def\ee{\end{equation}}
\def\ba{\begin{eqnarray}}
\def\ea{\end{eqnarray}}

%\preprint{DOE/ER/40561-50}

\title{Numerical study of a three-dimensional generalized stadium billiard}
\author{Thomas Papenbrock}
\address{Institute for Nuclear Theory, Department of Physics, 
University of Washington, Seattle, WA 98195, USA}
\maketitle
\begin{abstract}
We study a generalized three-dimensional stadium billiard and present strong
numerical evidence that this system is completely chaotic. In this convex
billiard chaos is generated by the defocusing mechanism. The construction of
this billiard uses cylindrical components as the focusing elements and thereby
differs from the recent approach pioneered by Bunimovich and Rehacek
\cite{BuRe96}. We investigate the stability of lower-dimensional invariant
manifolds and discuss bouncing ball modes.
\end{abstract}
Billiards are simple yet nontrivial models of classically chaotic systems.  Of
particular importance are Sinai's billiard \cite{SinaiB} and Bunimovich's
stadium \cite{BunimS} since they are proven to be completely chaotic and
ergodic. These two billiards exhibit two different chaos generating mechanisms,
namely dispersion and defocusing. Upon scattering off an dispersing boundary
element nearby trajectories diverge, and consecutive collisions with dispersing
elements lead to an increasing divergence. In focusing billiards nearby
trajectories converge after a collision with the focusing boundary element. It
is only after the trajectories pass through the focusing point that they start
to diverge. Provided the free flight (including reflections at neutral boundary
elements) is sufficiently long, the focusing may be over compensated by the
divergence and result in a defocusing. It is important to note that a weak
focusing requires a long free flight until defocusing results.  While
dispersing billiards like Sinai's may easily be generalized to more than two
dimensions, it was not until recently that completely chaotic
higher-dimensional focusing billiards were constructed. Bunimovich and Rehacek
proved that spherical caps attached to three-dimensional billiards with neutral
boundary elements may be chaotic and ergodic \cite{BuRe96}. The main difficulty
to overcome was the weak focusing in directions transverse to the plane that is
defined by consecutive scatterings of a trajectory with a spherical cap.
Bunimovich and Rehacek solved this problem by putting certain conditions on the
size and distance of spherical caps. This ensures that any focusing eventually
turns into defocusing. However, numerical studies showed that these conditions
may be relaxed \cite{BCG96}, and that the mechanism of defocusing also works 
beyond three dimensions \cite{BuRe98}. 

In this work we want to study a different construction of chaotic 
focusing billiards in three dimensions. The basic idea is as follows. 
Let us use cylindrical instead of spherical components as the focusing boundary
elements.  A cylindrical element focuses in two-dimensional planes 
perpendicular to the cylinder axis and is neutral in the directions of the 
axis. This avoids the problem caused by the weak focusing in spherical caps. 
The generation of high-dimensional chaos requires, however, more than one 
cylindrical boundary element, and their axes must be differently oriented. 
As an example we study numerically a three-dimensional generalization of the 
stadium billiard. Its construction is related to the 
construction of geodesic flows on high-dimensional (boundaryless) manifolds 
that are products of lower-dimensional manifolds \cite{BrunsGerber94}. 
Due to its use of cylindrical focusing elements it is also related to the
billiard model of a self-bound three-body system \cite{PP}. Unlike the
latter, the generalized three dimensional stadium can easily be realized 
experimentally. This is of particular interest with view on recent 
experiments that investigate wave chaos in three-dimensional elasto-mechanical 
systems \cite{Weaver,NBI} and microwave cavities \cite{Deus,Richter,Kuhl}.

Besides these possible applications the results presented in this work are also
of interest to further fields in physics. We recall that billiards are widely
being studied in the field of quantum chaos (for a review see, e.g.  reference
\cite{GMW}), and play an important role in extending this field to systems with
more than two degrees of freedom \cite{BCG96,PP,Primack,Prosen}. They are also
central to the investigation of three-dimensional wave chaos in resonant
optical cavities \cite{Nockel}.

\begin{figure}
  \begin{center}
    \leavevmode
    \parbox{0.4\textwidth}
           {\psfig{file=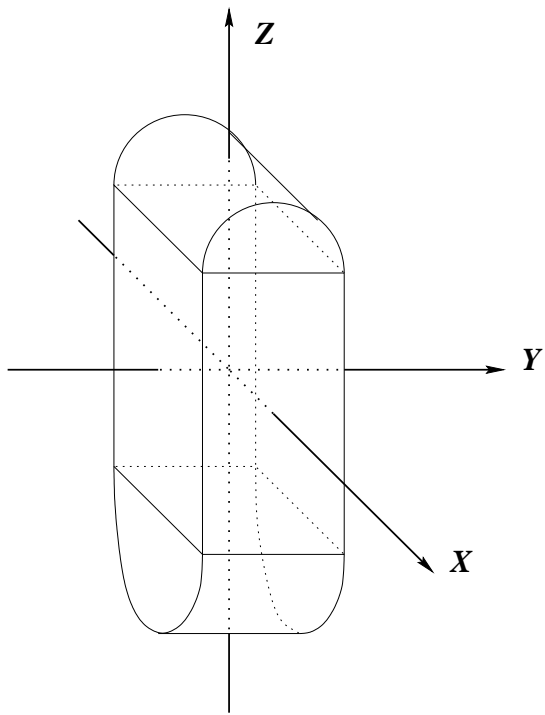,width=0.4\textwidth,angle=270}}
  \end{center}
\protect\caption{Three-dimensional generalization of the stadium billiard}
\label{fig1}
\end{figure}

Let us consider the three-dimensional billiard depicted in figure~\ref{fig1}.
We denote the radii of the lower and upper half-cylinder as $r_1$ and $r_2$,
respectively, and the distance between these half-cylinders as $2a$. This
billiard can be viewed as a generalization of the two-dimensional stadium to
three dimensions. Its sections with planes normal to the $x$- and $y$- axis are
partly desymmetrized two-dimensional stadia. One might expect that this system
displays high-dimensional chaos, since \cite{BunimQuote} ``In many-dimensional
billiards with chaotic behavior the local instability has to be in all its
two-dimensional sections''. Note, that the upper and lower half-cylinder
defocuses in the planes normal to the $x$-axis and the $y$-axis,
respectively. Note further that this property persists even when the distance
between the two half-cylinders approaches $2a=0$.

In what follows let us fix $r_1=r_2=r$ and $a=0$. To compute the Lyapunov
spectrum we draw $10^4$ uniformly distributed phase space points at random,
fixing the velocity $|\vec{v}|=1$. We follow the time evolution for each of
these phase space points for about $5\times 10^5$ bounces off the boundary and
compute the Lyapunov spectrum from the tangent map \cite{Dellago,Sieber}.  For
the time evolution we note that the particle moves freely inside the billiard
and undergoes specular reflections upon collisions with the boundary. Let
$\vec{v}$ and $\vec{v}'$ denote the velocity immediately before and after a
collision with the boundary, respectively. One has
$\vec{v}'=\vec{v}-2(\vec{v}\cdot\vec{n})\vec{n}$, where $\vec{n}$ is the unit
normal vector of the boundary at the collision point. The tangent map is a
product of maps corresponding to free flights and to reflections. It governs
the time evolution of infinitesimally small deviations from the trajectory. Let
$(\vec{x},\vec{v})$ and $(\vec{x}',\vec{v}')$ denote initial and final phase
space points of a free flight, respectively. The corresponding tangent map has
elements
\ba
\frac{\partial x_j'}{\partial x_k}=\delta_{jk}, \quad  
\frac{\partial x_j'}{\partial v_k}=t\,\delta_{jk}, \quad
\frac{\partial v_j'}{\partial x_k}= 0, \quad  
\frac{\partial v_j'}{\partial v_k}=\delta_{jk}, \qquad j,k=1,2,3.   
\nonumber
\ea
To describe a reflection at the cylindrical boundary element of radius $r$ we 
choose coordinates such that the cylinder axis is parallel to the $z$-axis.
Let $(\vec{x},\vec{v})$ and $(\vec{x}',\vec{v}')$ denote phase space points
immediately before and after a reflection, respectively. The corresponding
tangent map has elements 
\ba
\frac{\partial x_j'}{\partial x_k}&=&\delta_{jk}-2n_jn_k, \quad
\frac{\partial x_j'}{\partial v_k} = 0,\qquad j,k=1,2,3, \nonumber\\
\frac{\partial v_j'}{\partial v_k}&=&\delta_{jk}-2n_jn_k, \quad
\frac{\partial v_j'}{\partial x_3} = 0, \quad 
\frac{\partial v_3'}{\partial x_k} = 0, \qquad j,k=1,2,3, \nonumber\\
\frac{\partial v_j'}{\partial x_k}&=&-{2\over r}
\left(w_n\delta_{jk}+n_jw_k-n_kw_j-{w^2\over w_n}n_jn_k\right), \qquad j,k=1,2,
\nonumber
\ea 
where $\vec{n}$ is the outwards pointing unit normal vector at the boundary, 
$\vec{w}=(v_1,v_2)$ the velocity in the plane normal to the $z$-axis and 
its normal component $w_n=\vec{n}\cdot\vec{w}$. A reflection at the flat 
parts of the boundary may be described using the equations above after 
taking $r\to\infty$. 

In a conservative system with three degrees of freedom the Lyapunov exponents
come in pairs $(\lambda_j,\lambda_{-j}), j=1,2,3$ with
$\lambda_1\ge\lambda_2\ge\lambda_3=0$ and $\lambda_j+\lambda_{-j}=0$. Our
numerical results (mean values, variances, maximal and minimal values) are
listed in table~\ref{tab1}. We observe two positive Lyapunov exponents, thus
indicating that truly high-dimensional chaos has developed. Note that all of
the followed trajectories have positive Lyapunov exponents $\lambda_1$ and
$\lambda_2$. We checked our numerical results by comparing forward with
backward evolution and by using the alternative method pioneered by Benettin
{\it et al.}  \cite{Benettin}. Within our numerical accuracy we find one pair
of vanishing Lyapunov exponents and confirmed that the sum of conjugated
exponents vanishes. We repeated the computation for a larger ensemble of $10^5$
trajectories and a shorter time evolution of about $5\times 10^4$ bounces off
the boundary. This computation reproduced the mean values of table~\ref{tab1}
but the distributions were broader. This is due to the shorter time evolution
which leads to somewhat less converged Lyapunov exponents. Again, no single
stable trajectory was found. Therefore, our numerical results strongly suggests
that the system under consideration is completely chaotic.
 
It is interesting to investigate the stability of lower-dimensional invariant
manifolds. Such manifolds exist in systems with discrete symmetries and in
rotationally invariant many-body systems composed of identical particles
\cite{PapSel}, though they need not necessarily be connected to a discrete
symmetry \cite{Prosen}.  Since their stability properties may deviate
considerably from the system average, it is important to investigate them more
closely. In what follows let us consider two lower-dimensional invariant
manifolds, namely (i) $y=0, p_y=0$ and (ii) $y=\pm r/\sqrt{2}, p_y=0$ or
$z=r/\sqrt{2}, p_z=0$. Note that manifold (i) is a symmetry plane of the
billiard whereas manifold (ii) is a less trivial example of a low-dimensional
invariant manifold.  Note further that manifold (i) and manifold (ii) can be
viewed as a partly desymmetrized and a full two-dimensional stadium,
respectively. Though these manifolds are of measure zero in phase space, they
may exhibit special stability properties in transverse directions
\cite{PapSel,BCG96}. This behavior may cause wave function scarring upon
quantization \cite{Prosen,PSW,PP}. We start 1000 randomly drawn trajectories
inside each of the invariant manifolds and compute the Lyapunov spectrum by
following their time evolution for about $5\times 10^5$ collisions with the
boundary. One pair of Lyapunov exponents describes the stability in directions
transverse to the manifold while the remaining two pairs correspond to the
inside motion. The results listed in table~\ref{tab2} show that both invariant
manifolds are unstable inside and in the transverse direction.  A comparison
with table~\ref{tab1} shows that the local instability close to the invariant
manifolds deviates from the average instability inside the billiard. This is
not surprising since stability properties of invariant sets like periodic
orbits or low-dimensional manifolds fluctuate around the system average. In
low-dimensional open systems such a behavior may have considerable influence on
quantum transport \cite{Kaplan}. Our results hint at a generalization of these
observations to three-dimensions.  We note that the Lyapunov exponent inside
each manifold agrees with the one reported for the corresponding
two-dimensional stadium billiard \cite{BenStad}.

We no turn to the more general case $a>0$ and $r_1\ne r_2$. To be definite we
fix $a=1,r_1=\sqrt{2},r_2=\sqrt{3}$ and compute the Lyapunov spectrum from
$10^4$ trajectories with uniformly distributed random initial conditions and a
time evolution of about $5\times 10^5$ bounces off the boundary. As before, we
do not find a single stable trajectory and both Lyapunov exponents are
positive, i.~e. $\lambda_1= 0.185\pm 0.001, \lambda_2= 0.157\pm 0.001$ in units
of $1/a$. This shows that truly high-dimensional chaos exists for these
parameter values, too.

Let us also discuss focusing billiards in more than three dimensions.
Bunimovich's and Rehacek's construction has successfully been used to create 
chaos in four-dimensional billiards \cite{BCG96}, and it works in higher 
dimensions as well \cite{BuRe98}. It has the advantage that a single spherical
cap attached to a billiard with otherwise flat boundaries may be sufficient to
render the system chaotic. This is different with the cylindrical elements used
in this work. While we do not see any principal argument that would prohibit
the generation of chaos in high-dimensional billiards by means of cylindrical
components (i.~e. such that are focusing in a two-dimensional plane only), it
requires certainly several of such boundary elements to generate the desired
degree of local instability. The billiard model of a self-bound interacting
many-body system \cite{TP99} might be a promising candidate for such a
scenario. However, more work is necessary for a better understanding of
focusing cylindrical boundary elements in high-dimensional systems.

Finally, we want to comment on the role of bouncing ball orbits in the
three-dimensional generalized stadium billiard. For $a > 0$ there is a infinite
number of families of bouncing ball orbits, and the situation is similar to the
case of the three-dimensional Sinai billiard.  Theoretical \cite{Primack} and
experimental \cite{Richter} studies of this billiard show that the bouncing
ball modes \cite{bb2d} dominate the length spectrum (i.e. the Fourier transform
of the fluctuating part of the spectral density). In three dimensions, the
amplitude of each bouncing ball mode is enhanced by $O(k)$ ($k$ being the wave
vector) when compared with the amplitude of an unstable periodic orbit; an
infinite number of bouncing ball modes with different length thus dominates the
length spectrum practically at all length. This makes the semi-classical
analysis of level spectra in terms of periodic orbits a difficult task. The
situation is, however, different for $a=0$ and $r_1=r_2=r$.  In this case,
there are only two families of bouncing ball orbits having equal length
$4r$. Thus, the billiard considered in this work is a promising candidate for
further experimental and theoretical investigations of wave chaotic phenomena
in three dimensions.

In summary, we have studied a generalized three-dimensional stadium billiard
that is chaotic due to the defocusing mechanism. The construction uses
cylindrical components as focusing boundary elements and thereby differs from
the one proposed by Bunimovich and Rehacek. We presented strong numerical
evidence that the considered system displays hard chaos. In particular, we
found two positive Lyapunov exponents and confirmed the instability of
lower-dimensional invariant manifolds.

This work was supported by the Dept. of Energy under Grant DE-FG-06-90ER40561.
I thank L. A. Bunimovich for bringing ref. \cite{BrunsGerber94} to my attention
and H. Rehfeld for useful discussions on bouncing ball modes.

\begin{table}
\begin{tabular}{|c||d|d|d|d|}
$j$ &$\lambda_j$&$\Delta\lambda_j$&$\lambda_{min}^{(j)}$&$\lambda_{max}^{(j)}$ 
\\\hline
1   & 0.364     & 0.001           & 0.347               & 0.368           \\
2   & 0.330     & 0.001           & 0.314               & 0.334           \\
\end{tabular}
\protect\caption{Results on Lyapunov exponents (mean values, 
variances, minimal and maximal values) obtained from an ensemble of $10^4$ 
runs. All quantities are given in units of $1/r$.}  
\label{tab1}
\end{table}

\begin{table}
\begin{tabular}{|c||d|d|}
manifold & $\lambda_{||}$    & $\lambda_{\perp}$\\\hline
(i)      &  0.430$\pm$0.003  & 0.305$\pm$0.002  \\
(ii)     &  0.391$\pm$0.004  & 0.362$\pm$0.004  \\
\end{tabular}
\protect\caption{Results on Lyapunov exponents (mean values and variances) 
for invariant manifolds, obtained from ensembles of $10^3$ runs. 
$\lambda_{||}$ and $\lambda_{\perp}$ denote the Lyapunov exponents inside
and transverse to the corresponding invariant manifold. All 
quantities are given in units of $1/r$.}  
\label{tab2}
\end{table}

\end{document}